\documentclass[useAMS,usenatbib]{mn2e}
\usepackage{epsf,psfrag,graphicx}
\usepackage[usenames]{color}

\newcommand{\pc}{\,{\rm pc}}
\newcommand{\Mv}{\hbox{$M_{\rm vir}$}}
\newcommand{\kpc}{\,{\rm kpc}}

\def\gtrsim{\lower.5ex\hbox{$\; \buildrel > \over \sim \;$}}
\def\MN{MN}
\def\CD{ART}
\def\cmmc{\,{\rm cm}^{-3}}
\def\msun{{\rm M_\odot}}

\title[Co-planarity of cold stream delivered satellites]{The co-planarity of
satellite galaxies delivered by randomly aligned cold mode accretion streams}

\author[Tobias Goerdt et al.]
{\parbox[t]{\textwidth}{Tobias Goerdt$^1$\thanks{tobias.goerdt@univie.ac.at},
Andreas Burkert$^{2,3}$ and Daniel Ceverino$^{4,5}$}\\  \vspace*{3pt} \\
$^1$Institut f\"ur Astrophysik, Universit\"at Wien, T\"urkenschanzstra{\ss}e
17, 1180 Wien, \"Osterreich\\
$^2$Universit\"ats-Sternwarte M\"unchen, Scheinerstra{\ss}e 1, 81679 M\"unchen,
Deutschland\\
$^3$Max Planck Fellow, Max-Planck-Institut f\"ur extraterrestrische Physik,
Gie{\ss}enbachstra{\ss}e, 85748 Garching, Deutschland\\
$^4$Centro de Astrobiolog{\'i}a (INTA-CSIC), Ctra de Torrej{\'o}n a Ajalvir, km
4, 28850 Torrej{\'o}n de Ardoz, Madrid, Espa\~na\\
$^5$Astro-UAM, Universidad Aut\'onoma de Madrid, Unidad Asociada CSIC, 28049
Madrid, Espa\~na\\}

\date{Draft version \today}
\pagerange{\pageref{firstpage}--\pageref{lastpage}}
\pubyear{2015}

\begin{document}

\maketitle

\label{firstpage}

\begin{abstract}
Recent observations have shown that the majority of the Andromeda galaxy's
satellites are aligned in a thin plane. On the theoretical side it has been
proposed that galaxies acquire their gas via cold streams. In addition,
numerical simulations show that the same streams also deliver gas clumps which
could potentially develop into satellite galaxies. Assuming that cold streams
are a major source of satellite systems around galaxies we calculate the
probabilities in two different models to find a certain fraction of satellites
within a thin plane around the central galaxy of the host halo with and without
having the same sense of rotation. Using simple geometrical considerations and
adopting a random orientation of the streams we demonstrate that the vast thin
disk of satellites detected around Andromeda can naturally be explained within
this framework. In fact, without any satellite scattering, two streams would
lead to too many satellites in the thin plane, compared with the observations.
Three streams reproduce the observations very well. Natural implications from
our model are that all massive galaxies should have a thin plane of satellites
and that the satellites should naturally distribute themselves not only into a
single plane but into several inclined ones. We estimate the effect of
additional satellites accreted from random directions and find it to be of
minor relevance for a mild inflow of satellites from random directions.
\end{abstract}


\section{Introduction}
\label{sec:intro}

Our understanding of how galaxies form has changed substantially in recent
years. A decade ago it was thought \citep{blumenthal, rees, silk, white} that
galaxies collect their baryons through diffuse gas, spherically symmetrically
falling into dark matter haloes and being shock-heated as it hits the gas
residing in the haloes, the so-called hot mode accretion. Whether the gas
eventually settles into the equatorial plane, forming a galactic disk was
depending on the mass of the dark halo. Below a critical mass, the gas could
cool efficiently, forming a disk galaxy, while for larger masses the cooling
time would be longer than the Hubble time, leading to structures that resemble
galaxy clusters with a large baryon fraction in the hot, diffuse intergalactic
gas component. Recent theoretical work and simulations \citep{fardal, bd03,
keresa, keresb, db06, ocvirk, DekelA_09a, dekel13} however seem to indicate
that at high redshift $(z \gtrsim 2)$, galaxies acquire their baryons primarily
via cold streams of relatively dense and pristine gas with temperatures around
$10^4$ K that penetrate through the diffuse shock-heated medium, the so-called
cold mode accretion. These streams peak in activity around redshift 3. Having
reached the inner parts of the host halo they will eventually form a dense,
unstable, turbulent disc where rapid star formation is triggered \citep{oscara,
oscarb, DekelA_09b, cd, andi, c12, marcello, genel12, genzel11, dekel13, cip}.

$N$-body simulations suggest that about half the mass in dark-matter haloes is
built-up smoothly, suggesting that the baryons are also accreted
semi-continuously as the galaxies grow \citep{genel10}. Hydrodynamical
cosmological simulations also show rather smooth gas accretion, including
mini-minor mergers with mass ratios smaller than 1:10, that brings in about two
thirds of the mass \citep{DekelA_09a}. The massive, clumpy and star-forming
discs observed at $z\sim 2$ \citep{genzel08, genel, foerster2, foerster3} may
have been formed primarily via the smooth and steady accretion provided by the
cold streams, with a smaller contribution by major merger events \citep{oscara,
cd}.

Attempts to directly observe cold accretion streams are ongoing: \citet{mich}
used cosmological hydrodynamical \textsc{amr} simulations to predict the
characteristics of Ly$\alpha$ emission from the cold gas streams. The
Ly$\alpha$ luminosity in their simulations is powered by the release of
gravitational energy as the gas is flowing inwards with a rather constant
velocity. The simulated Ly$\alpha$-blobs (LABs) are similar in many ways to the
observed LABs. Some of the observed LABs may thus be regarded as direct
detections of the cold streams that drove galaxy evolution at high $z$.
Observations \citep{rauch, erb} and new \textsc{amr} simulations incorporating
radiative transfer support this model \citep{joki}. \citet{mich2} made
theoretical predictions about the likelihood of observing these streams in
absorption which have very recently really been observed \citep{bouche2}.
\citet{mich5} looked at the amount of inflow -- the mass accretion rate -- both
as a function of radius, mass and redshift for the three constituents gas,
stars and dark matter.

\citet{danovich} find that at the few virial radii vicinity of the galaxy, the
streams tend to be confined to a stream plane, and embedded in a flat pancake
that carries $\sim 20\%$ of the influx. There are on average three significant
streams, of which one typically carries more than half the mass inflow. Given
the fact that galaxies grow by cold stream accretion, the question arises
whether some observational signatures of this phase are still detectable in
present-day galaxies. \citet{mich4} analysed the velocity of the accretion
along the streams.

\begin{figure*}
\begin{center}
\includegraphics[width=17.73cm]{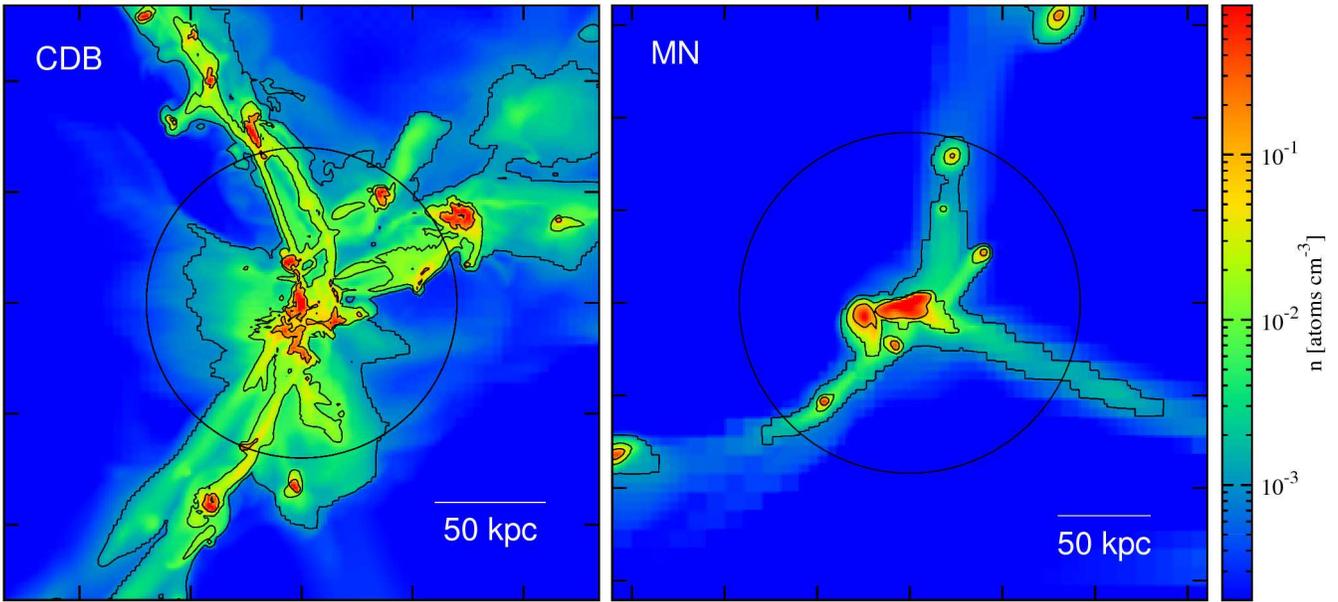}
\end{center}
\caption{Gas density in simulated galaxies from {\CD} and {\MN}. The colours
refer to the maximum density along the line of sight. The contours mark $n =
0.1$, 0.01 and $0.001\cmmc$, respectively. The circles show the virial radii.
Left: a typical {\CD} galaxy (resolution $70\pc$) at $z=2.3$, with $\Mv = 3.5
\times 10^{11}\msun$ at $z=2.3$. Right: one of the {\MN} galaxies (resolution
$1\kpc$) at $z=2.5$, with $\Mv = 10^{12}\msun$ at $z=2.5$. In both cases, the
inflow is dominated by three cold narrow streams that are partly clumpy. The
density in the streams is $n=0.003-0.1 \cmmc$, with the clump cores reaching $n
\sim 1\cmmc$. An example for a stream without clumps just consisting of smooth
gas is seen in the lower right hand corner of the right hand panel. There is
nothing peculiar about such a 'clump-less' stream.}
\label{fig:denmap}
\end{figure*}

\citet{noam1,noam2} looked at the preferred direction and at the distribution
of satellite galaxies in the Local Group. They used $N$-body dark matter only
simulations. They find that the satellites within the Local Group should
preferentially lie in a plane.

\citet{ibata} showed in a seminal observational paper the existence of a planar
subgroup of satellites in the Andromeda galaxy (M 31), comprising about half of
the population as observed and described by \citet{conn}. The structure is at
least 200 kilo-parsec in radius, but also extremely thin, with a perpendicular
scatter of less than 12.6 kilo-parsec. Radial velocity measurements reveal that
the satellites in this structure have the same sense of rotation about their
host. This shows conclusively that substantial numbers of dwarf satellite
galaxies share the same dynamical orbital properties and direction of angular
momentum. Intriguingly, the plane they identify is approximately aligned with
the pole of the Milky Way's disk and with the vector between the Milky Way and
Andromeda. \citet{ibata2} stressed the importance of reproducing all of the
observed properties of the plane, especially the fact that most of the
satellites are co-rotating, when comparing thy numerical simulations.
\citet{ibata3} found observational evidences that thin planes of satellites
might be a common phenomenon amongst galaxies in the universe.

Following up on this \citet{bowden} argued that a thin satellite disc can
persist over cosmological times if and only if it lies in the planes
perpendicular to the long or short axis of a triaxial halo, or in the
equatorial or polar planes of a spheroidal halo. In any other orientation, the
disc thickness would double on $\sim 5$ Gyr timescales and so must have been
born with what they call 'an implausibly small vertical scale-height'.

Several scenarios have been proposed in order to explain the origin of planar
satellite systems. Dwarf galaxies might be accreted in groups \citep{donghia08,
li08}. Alternatively, the disk of satellites might be the tidal debris of a
major merger with a gas-rich galaxy with the tidal arms condensing into tidal
dwarf galaxies \citep{wetzstein07, bournaud08, pawlowski11, hammer}. This
scenario is however in conflict with the observational evidence of substantial
amounts of dark matter, dominating the kinematics of dwarf spheroidals
\citep[see however][]{kroupa97, pawlowski12}. Interestingly, coherently
rotating, quasi-planar distributions of satellites have also been found in
cold-dark-matter simulations \citep{lovell11, keller12, bb14, gillet}
\citep[see][for an alternative view]{pawlowskiemail, gerhard}. \citet{andrea}
found evidence in numerical simulation for the creation of polar ring galaxies
by cold stream accretion. \citet{temple1} showed that galaxy pairs align with
galactic filaments and \citet{temple2} found alignement between the angular
distribution of satellite galaxies around the isolated primary galaxies in
filaments and the direction of the filaments where those primaries are located
in SDSS observations. This indicates that satellite great planes might be a
natural result of how galaxies accrete material and substructure from the
cosmic web. It is this idea, that is the motivation for our paper. We
demonstrate that thin planes of substructure, consistent with the observations
of \citet{ibata} are in fact expected if galaxies are fed by cold, satellite
loaded streams. In section \ref{sec:calc} we present two simple geometrical
models to analyse the probability of generating a satellite disk with given
thickness and satellite fraction, for various numbers of randomly oriented cold
streams. Section \ref{sec:conc} presents the conclusions.

\section{Calculations}
\label{sec:calc}

It has been proposed that gas enters a host halo via smooth accretion streams
\citep{DekelA_09a}. For our purposes more interestingly, dark matter haloes
filled with gas that could potentially form stars and end up as satellite
galaxies are seen in simulations to lie on top of those smooth accretion
streams (e.g. figure 2 of \citet{DekelA_09a} shows gas clumps on top of gas
streams, which are identified by \citet{mich3} to be gas clumps surrounded by
dark matter haloes). In figure \ref{fig:denmap} we show maps from
hydrodynamical \textsc{amr} simulations of high redshift $(z \sim 2.5)$, high
mass $(M_{\rm vir} > 10^{11}$ M$_\odot)$ galaxies from two suites of cosmological
simulations: the Horizon-MareNostrum simulation \citep[hereafter
{\MN},][]{ocvirk} and the {\CD} suite of simulations \citep[Adaptive Refinement
Tree,][]{cak, cd, c12, dekel13, c14a}. The circles indicate the virial radii.
Clearly visible are three cold streams, funnelling gas into the centre of the
galaxy. One can also see several gas clumps, embedded in the streams. Those are
smaller galaxies which are entering the host halo via the streams. These
smaller structures could merge with the central galaxy or, as we will assume
below, end up as satellites, orbiting their host. Note that there are no clumps
outside a stream, indicating that the build-up of a satellite system is stream
driven \citep[see e.g:][]{tormen, onuora, noam1, DekelA_09a}. Note further that
the clumps are carried in the innermost parts of the streams which are
substantially narrower than 13 kpc (i.e. the width of the plane of satellites
of M31).

Suppose now that galaxies and their satellite systems form in the focal point of
a number $m$ of randomly oriented cold streams. What is then the probability
that a certain fraction of satellites lies within a thin plane around the
central galaxy of the host halo? \citet{ibata} found through observations that
15 out of Andromeda galaxy's 27 satellites are within a plane with 200 kpc
radius and 12.6 kpc vertical rms scatter. As the radius (total extent) and
vertical thickness (rms scatter) are defined in a different way these values
cannot be used directly in order to determine the opening angle of the disk. We
therefore analysed again the 3-D coordinates of the 15 satellite galaxies
selected by \citet[][see figure 10 of Conn et al. 2012]{ibata} and computed the
rms radius $R$ instead of the maximum radius to be able to compare it to the
rms thickness $D$ (12.6 kpc) of the disk. The rms radius turns out to be 187
kpc. From that we computed the opening angle atan$(D/R)$ which is 3.8 degree.
\citet{ibata} excluded two out of those 15 subhaloes since those are
counter-rotating. Since our first analysis using radial orbits does not take
the orientation of the rotation into account we will compare this analysis to
the ratio $15 / 27 \simeq 56\, \%$. Our second analysis, deploying planar
orbits does take the orientation of the rotation into account therefore we will
compare that analysis to the ratio $13 / 27 \simeq 48\, \%$.

In order to work out the likelihood of having of order half of the satellites
in a plane as thin as 3.8 degree we have performed two sets of Monte Carlo
experiments, drawing randomly orientated streams and calculating the
probabilities of having a given ratio of subhaloes within a plane of certain
thickness. Underlying our calculations are the following assumptions: (1) A
vast majority of incoming subhaloes enter the host halo through cold streams,
as seen in the arrangement of the gas clumps in figure 2 of \citet{DekelA_09a}.
The effect of having additional subhaloes entering from random
directions\footnote{e.g. accretion of satellites at cosmic times at which there
is no cold stream activity, i.e. outside $z \sim 1-4$.} will be discussed in
depth in section \ref{sec:addsph}. (2) Some of the gas clumps carried by the
cold streams end up as satellite galaxies orbiting the central galaxy today.
(3) Any host halo has between two and seven streams. (4) The streams themselves
are randomly distributed over the whole sky as seen from the centre of the host
halo. We are aware of studies which present evidence that already the streams
seem to lie in a single plane \citep{DekelA_09a, danovich}. It is therefore a
conservative assumption to distribute the streams randomly with equal solid
angles on the sky carrying equal probabilities of having a stream. (5) The
orientation of the cold streams does not change with respect to the host halo
during the period of accreting satellites. Both of our models are built upon
those five assumptions.

\begin{figure*}
\psfrag{Psi [degree]}[B][B][1][0]{$\Psi$ [degree]}
\psfrag{cumulative distribution P}[B][B][1][0]{cumulative distribution $P$}
\psfrag{Ibata et al.}[Br][Br][1][0]{Ibata et al.}
\psfrag{100 = 2 / 2}[Br][Br][1][0]{100 \% = 2 / 2}
\psfrag{ 67 = 2 / 3}[Br][Br][1][0]{67 \% = 2 / 3}
\psfrag{100 = 3 / 3}[Br][Br][1][0]{100 \% = 3 / 3}
\psfrag{ 50 = 2 / 4}[Br][Br][1][0]{50 \% = 2 / 4}
\psfrag{ 75 = 3 / 4}[Br][Br][1][0]{75 \% = 3 / 4}
\psfrag{100 = 4 / 4}[Br][Br][1][0]{100 \% = 4 / 4}
\psfrag{ 40 = 2 / 5}[Br][Br][1][0]{40 \% = 2 / 5}
\psfrag{ 60 = 3 / 5}[Br][Br][1][0]{60 \% = 3 / 5}
\psfrag{ 80 = 4 / 5}[Br][Br][1][0]{80 \% = 4 / 5}
\psfrag{100 = 5 / 5}[Br][Br][1][0]{100 \% = 5 / 5}
\psfrag{ 33 = 2 / 6}[Br][Br][1][0]{33 \% = 2 / 6}
\psfrag{ 50 = 3 / 6}[Br][Br][1][0]{50 \% = 3 / 6}
\psfrag{ 67 = 4 / 6}[Br][Br][1][0]{67 \% = 4 / 6}
\psfrag{ 83 = 5 / 6}[Br][Br][1][0]{83 \% = 5 / 6}
\psfrag{100 = 6 / 6}[Br][Br][1][0]{100 \% = 6 / 6}
\psfrag{ 29 = 2 / 7}[Br][Br][1][0]{29 \% = 2 / 7}
\psfrag{ 43 = 3 / 7}[Br][Br][1][0]{43 \% = 3 / 7}
\psfrag{ 57 = 4 / 7}[Br][Br][1][0]{57 \% = 4 / 7}
\psfrag{ 71 = 5 / 7}[Br][Br][1][0]{71 \% = 5 / 7}
\psfrag{ 86 = 6 / 7}[Br][Br][1][0]{86 \% = 6 / 7}
\psfrag{100 = 7 / 7}[Br][Br][1][0]{100 \% = 7 / 7}
\psfrag{2 streams}[Br][Br][1][0]{2 streams}
\psfrag{3 streams}[Br][Br][1][0]{3 streams}
\psfrag{4 streams}[Br][Br][1][0]{4 streams}
\psfrag{5 streams}[Br][Br][1][0]{5 streams}
\psfrag{6 streams}[Br][Br][1][0]{6 streams}
\psfrag{7 streams}[Br][Br][1][0]{7 streams}
\psfrag{radial orbits}[Br][Br][1][0]{radial orbits}
\psfrag{$1$}[B][B][1][0]{$1$}
\psfrag{$0.9$}[B][B][1][0]{$0.9$}
\psfrag{$0.8$}[B][B][1][0]{$0.8$}
\psfrag{$0.7$}[B][B][1][0]{$0.7$}
\psfrag{$0.6$}[B][B][1][0]{$0.6$}
\psfrag{$0.5$}[B][B][1][0]{$0.5$}
\psfrag{$0.4$}[B][B][1][0]{$0.4$}
\psfrag{$0.3$}[B][B][1][0]{$0.3$}
\psfrag{$0.2$}[B][B][1][0]{$0.2$}
\psfrag{$0.1$}[B][B][1][0]{$0.1$}
\psfrag{$5$}[B][B][1][0]{$5$}
\psfrag{$10$}[B][B][1][0]{$10$}
\psfrag{$15$}[B][B][1][0]{$15$}
\psfrag{$20$}[B][B][1][0]{$20$}
\psfrag{$25$}[B][B][1][0]{$25$}
\psfrag{$30$}[B][B][1][0]{$30$}
\psfrag{$35$}[B][B][1][0]{$35$}
\psfrag{$40$}[B][B][1][0]{$40$}
\psfrag{$45$}[B][B][1][0]{$45$}
\begin{center}
\includegraphics[width=17.73cm]{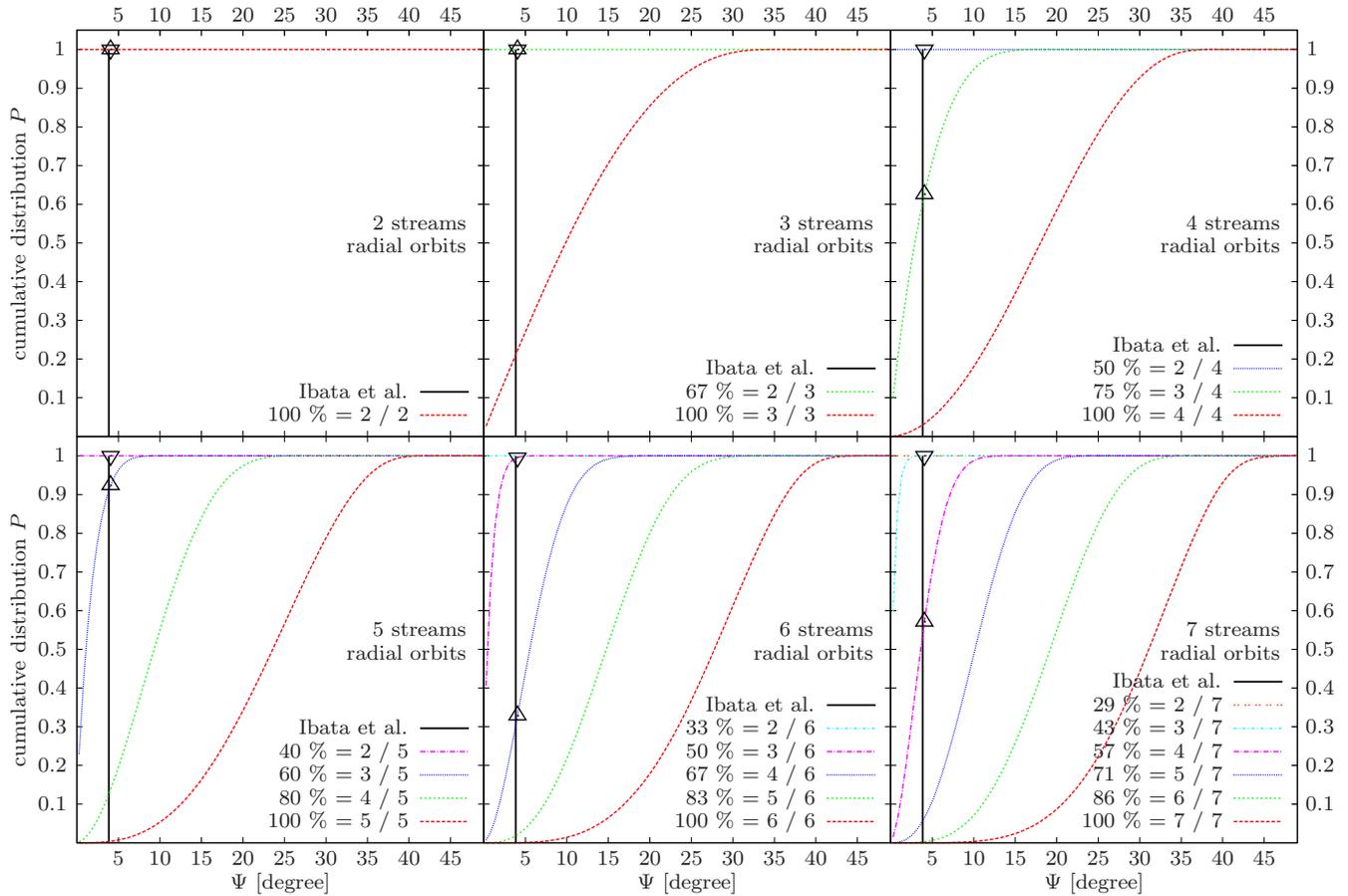}
\end{center}
\caption{Radial orbits model: The coloured dashed lines show the cumulative
distribution $P$ of having $n$ out of the total $m$ streams within an angle
$\Psi$. The solid black vertical line indicates the angle of 3.8 degree
estimated for Andromeda's thin satellite disk. The two triangles show the
intersection of this vertical line with the two curves which are the closest to
the observational value of 56\% (15 out of 27). They indicate the likelihood
that the observed fraction of streams lie in a single plane with at most the
observed angular deviation. In this model one would expect a thin disk of
satellites as observed for all different configuration of two to seven cold
streams.}
\label{fig:streams}
\end{figure*}

\subsection{Radial orbits model}
\label{sec:rom}
For this model, which deploys radial orbits, an additional four assumptions are
needed: (6) Each stream is loaded with an equal amount of subhaloes. (7) The
streams will hit the centre of the host halo directly head on, i.e. they have
no significant impact parameter with respect to the central galaxy. (8) The
subhaloes will initially stay on the same orbits as defined by their stream,
i.e. radial orbits parallel to the stream the subhalo is coming from. (9) These
orbits will eventually be circularised within a plane also parallel to this
stream.

Using these assumptions we have performed a series of Monte Carlo calculations.
We start a Monte Carlo calculation by fixing the total number $m$ of streams,
with $m$ in the range of two to seven. We then draw a large number $(10^7)$ of
sets of $m$ random stream orientations. As mentioned earlier, randomly oriented
means, that an equal solid angle of the sky as seen from the centre of the host
halo has an equal probability of carrying a stream. Also all streams must reach
the centre of the host halo directly and the position of one of the streams
does not have any influence on the position of the other streams. For each set
of streams we conduct for each number $n$ ($n \le m$) of streams out of the
total $m$ streams a grid search for the 'optimum' plane. 'Optimum' means that
the largest of $n$ angles between $n$ streams and the plane is minimal. For
this we rasterise the whole unit sphere centred on the host halo with points
equally distributed and never further apart than 0.2 degree. Equally
distributed means in this context that all points on the sphere are angularly
equidistant. The vectors connecting these points with the centre of the host
halo are the normal vectors to our plane candidates. A single subset of $n$
streams out of the total of $m$ streams gives a single optimal plane, but there
are different optimal planes for different values of $n$ within the same set of
$m$ total streams. In practice, the program runs through the following steps:
Given one of the planes from our grid search mentioned above, we calculate the
angles between each of the $m$ streams and the plane. The resulting $m$ angles
are sorted according to their value. The highest value denotes the opening
angle within which all $n = m$ streams lie in one plane. The second highest
value indicates the opening angle within which a number $n = m - 1$ streams
belong to one plane. We collect the values of the minimal angles $\Psi$ for
each set of streams and for each value of $n$. We then proceed to the next test
plane of our grid search.

\subsubsection{Results}
\label{sec:romresu}

In figure \ref{fig:streams} we show the cumulative distribution functions of
having $n$ out of $m$ streams close to a plane with a maximum angular deviation
of $\Psi$ (opening angle). These are the cumulative distribution functions of
the minimum angle $\Psi$ for a set of $n$ streams out of the total of $m$
streams. Each panel corresponds to a different total number $m$ and each of the
coloured dashed curves shows a different value of $n$. The vertical black line
at 3.8 degree indicates the observation of \citet{ibata}. The two triangles on
that line indicate the intersections with those predicted $n / m$ curves that
lie most closely to the fraction $15/27 \simeq 56\, \%$ of Andromeda's
satellites that have been found to belong to the thin plane. These triangles
therefore show upper and a lower values for our predicted probabilities. In the
first five columns of table \ref{tab:nstreams} we explicitly state the
positions of the triangles.

\begin{table}
\begin{center}
\setlength{\arrayrulewidth}{0.5mm}
\begin{tabular}{ccclll}
\hline
$m$ & $n_{\rm upper}$ & $n_{\rm lower}$ & $P_{\rm upper}$ & $P_{\rm lower}$ &
$P_{\rm planar}$ \\
\hline
2 & 2 & 2 & 1.0   & 1.0   & 1.0   \\
3 & 2 & 2 & 1.0   & 1.0   & 0.588 \\
4 & 3 & 2 & 1.0   & 0.622 & 0.243 \\
5 & 3 & 2 & 1.0   & 0.920 & 0.115 \\
6 & 4 & 3 & 0.996 & 0.317 & 0.078 \\
7 & 4 & 3 & 1.0   & 0.558 & 0.053 \\
\hline
\end{tabular}
\end{center}
\caption{Quoted are for all possible total numbers $m$ of streams $m$ the two
values for $n$ whose ratios are just above $(n_{\rm upper})$ and just below
$(n_{\rm lower})$ the observed ratio of $15 / 27 \simeq 56\, \%$ (radial orbits
model). In the next two columns the cumulative probability $P$ that $n$ out of
$m$ streams lie in a single plane with a deviation of less than the observed
3.8 degree is quoted (radial orbits model). The final column denotes the
probability $P_{\rm planar}$ of having $13 / 27 \simeq 48\, \%$ satellites within
a single thin plane and also having the same sense of rotation within the
planar orbits model.}
\label{tab:nstreams}
\end{table}

\begin{figure*}
\begin{center}
\includegraphics[width=17.73cm]{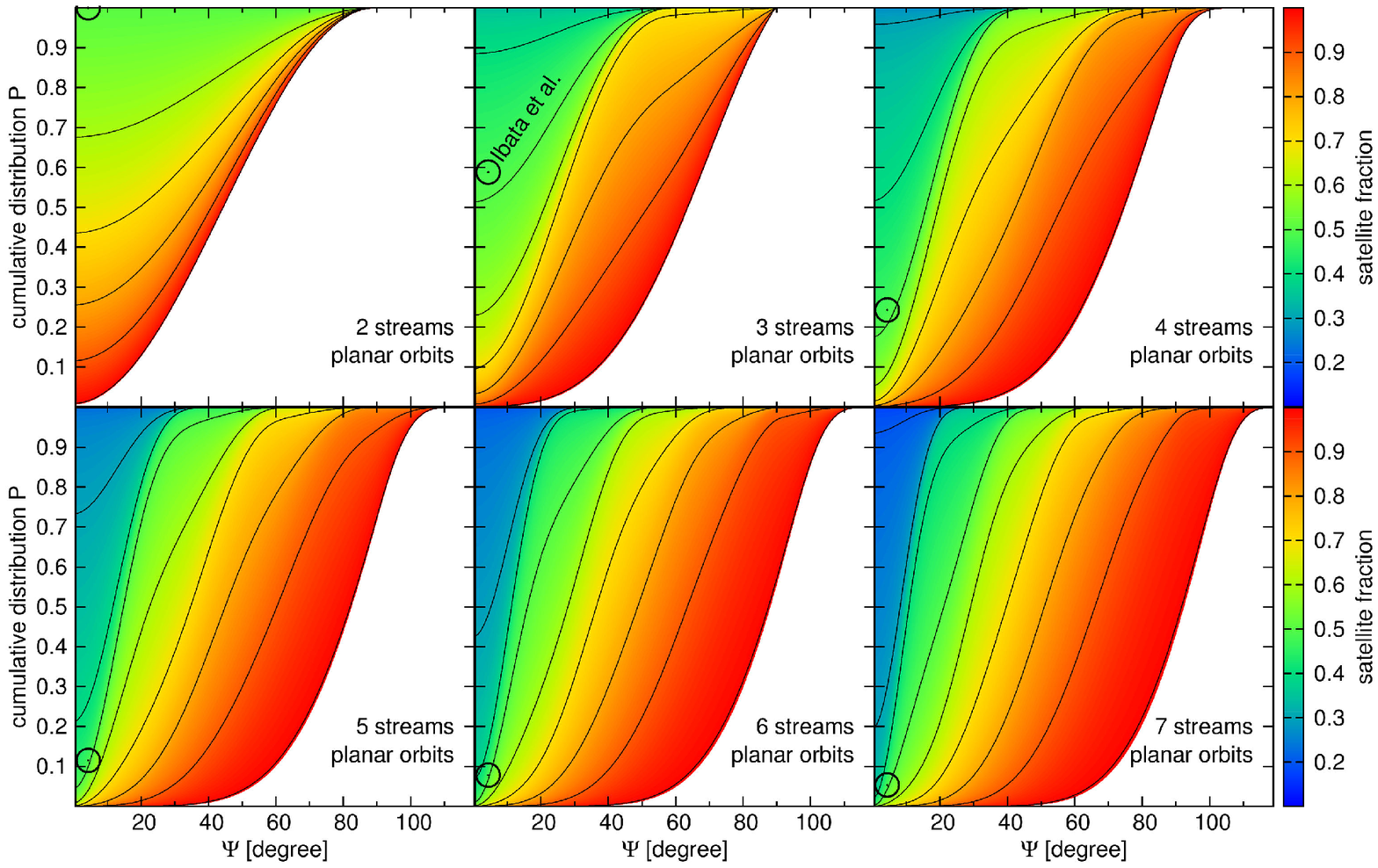}
\end{center}
\caption{Planar orbits model: Cumulative distribution $P$ of having the orbital
normal vectors of a certain satellite fraction within an angle $\Psi$. The
black contour lines are at satellite fractions of 0.2, 0.3, 0.4, 0.5, 0.6, 0.7,
0.8, 0.9 and 1.0, respectively. The panels indicate the likelihood of a given
fraction of satellites lying in a single thin plane and the same sense of
rotation. The black circles indicate the observational value of 48\% (13 out of
27) within 3.8 degree. In this model one would expect a thin disk of satellites
as observed in a two or three streams scenario which are the most common ones
in cosmological simulations.}
\label{fig:postreams}
\end{figure*}

Since we assumed that all the streams pass trough the galaxy centre, two
streams will always lie on a perfect plane. The '2 / $m$' lines in figure
\ref{fig:streams} therefore are always at a probability of $P=1.0$. For a three
stream scenario we would expect at least 67 \% of the satellites to lie within
one plane. Since two, three or four stream configurations are most commonly
seen in cosmological simulations \citep{danovich}, we can already say that
there is nothing surprising about the configuration seen around the Andromeda
galaxy. On the contrary, the fact that so little satellites lie on a thin plane
rules out a two stream or a three stream scenario for the Andromeda galaxy,
unless secular evolution or other processes move some of the satellites out of
that plane.

Since all heavy galaxies are fulled by cold streams it is a natural consequence
of this model that not only the Andromeda galaxy, but all galaxies having at
least Andromeda's mass should have a thin plane of satellites as Andromeda
does. Recent observational work \citep{ibata3, ibata4} points in this direction
\citep[See][however for an alternative view]{cautun}. The beauty of this model
is that it can also explain the two counter-rotating satellite galaxies which
belong to the plane of satellites. Other models have severe difficulties with
explaining those two.

\subsection{Planar orbits model}
\label{sec:pom}

Since \citet{ibata2} pointed out the importance of reproducing all of the
observed properties of the plane we decided to refine our model further,
allowing for planar, co-rotating orbits. For this second model we use the
common five assumptions from the beginning of section \ref{sec:calc}, but we
exchange the old assumptions 6 to 9 and replace them by three more realistic
ones: (6) The streams are loaded with a varying amount of subhaloes. Each
stream has the same probability of carrying many or only few satellites. (7)
The streams will not hit the centre of the host halo head on, but with a
significant impact parameter in a random direction perpendicular to the stream
itself. (8) The subhaloes will stay on planar orbits, in a way that the cross
product of the vector of the stream and the vector of the impact parameter
defines the plane of the orbit and the sense of the rotation within this plane.

Using these assumptions we have performed another large number $(10^7)$ of
Monte Carlo calculations. This second set of Monte Carlo simulations is very
similar to the first set mentioned in section \ref{sec:rom}: First we fix the
total number $m$ of streams, with $m$ being in the range of two to seven. We
then draw a large number of sets of $m$ random stream orientations. In this
model we also draw for each stream a random impact parameter direction. The
impact parameter direction is a unit vector which has to be perpendicular to
the stream orientation vector which is also a unit vector. The amplitude of the
impact parameter vector is always unity (the actual physical size of the impact
parameter is irrelevant to our problem, due to the expected circularisation of
the orbits) and its remaining direction angle (the overall vector has to be
perpendicular to the stream orientation vector) is uniformly distributed
between 0 and 2 $\pi$. Also for each stream we draw a satellite loading factor
in a way that each stream has the same satellite loading factor distribution
and the sum of the satellite loading factors of all $m$ streams always adds up
to one. This is done by drawing a random number uniformly distributed between 0
and 1 for each of the $m$ streams. These numbers are then renormalised so that
their sum equals to one. This approach might lead to streams which carry only a
single or even no clumps. There is nothing peculiar about such a stream
carrying only smooth gas but no clumps, it is still a perfectly valid stream.
Simulations show indeed that such streams exists. An example for such a stream
is the one plotted in the right hand panel of figure \ref{fig:denmap} (the
stream in the lower right hand corner). We now conduct a grid search for the
'optimum' plane the same way we did in our first model described in section
\ref{sec:rom}. Now we collect the values of the minimal angles $\Psi$ for each
possible combination of satellite loading factors.

\subsubsection{Results}
\label{sec:pomresu}

In figure \ref{fig:postreams} we plot the cumulative distribution functions of
having the indicated fraction of satellites in a plane with an opening angle
$\Psi$ and the same sense of rotation. Each panel corresponds to a different
total number $m$ of streams. The black circles indicate the observation of
Andromeda which has a fraction $13 / 27 \simeq 48\, \%$ of its satellites
within a thin plane with an opening angle of 3.8 degree and the same sense of
rotation \citep{ibata}. In the last column of table \ref{tab:nstreams} we
explicitly state the positions of the circles.

\begin{figure*}
\psfrag{cumulative distribution P}[B][B][1][0]{cumulative distribution $P$}
\psfrag{Ibata et al.}[Br][Br][1][0]{Ibata et al.}
\psfrag{2 streams}[Br][Br][1][0]{2 streams}
\psfrag{3 streams}[Br][Br][1][0]{3 streams}
\psfrag{4 streams}[Br][Br][1][0]{4 streams}
\psfrag{5 streams}[Br][Br][1][0]{5 streams}
\psfrag{6 streams}[Br][Br][1][0]{6 streams}
\psfrag{7 streams}[Br][Br][1][0]{7 streams}
\psfrag{radial orbits}[B][B][1][0]{radial orbits}
\psfrag{extended radial orbits}[Bl][Bl][1][0]{extended radial orbits}
\psfrag{planar orbits}[Bl][Bl][1][0]{planar orbits}
\psfrag{frandom}[B][B][1][0]{$f_{\rm random}$}
\psfrag{$1$}[B][B][1][0]{1}
\psfrag{$0.9$}[B][B][1][0]{0.9}
\psfrag{$0.8$}[B][B][1][0]{0.8}
\psfrag{$0.7$}[B][B][1][0]{0.7}
\psfrag{$0.6$}[B][B][1][0]{0.6}
\psfrag{$0.5$}[B][B][1][0]{0.5}
\psfrag{$0.4$}[B][B][1][0]{0.4}
\psfrag{$0.3$}[B][B][1][0]{0.3}
\psfrag{$0.2$}[B][B][1][0]{0.2}
\psfrag{$0.1$}[B][B][1][0]{0.1}
\psfrag{$0$}[B][B][1][0]{0}
\psfrag{$5$}[B][B][1][0]{5}
\psfrag{$10$}[B][B][1][0]{10}
\psfrag{$15$}[B][B][1][0]{15}
\psfrag{$20$}[B][B][1][0]{20}
\psfrag{$25$}[B][B][1][0]{25}
\psfrag{$30$}[B][B][1][0]{30}
\psfrag{$35$}[B][B][1][0]{35}
\psfrag{$40$}[B][B][1][0]{40}
\psfrag{$45$}[B][B][1][0]{45}
\begin{center}
\includegraphics[width=17.73cm]{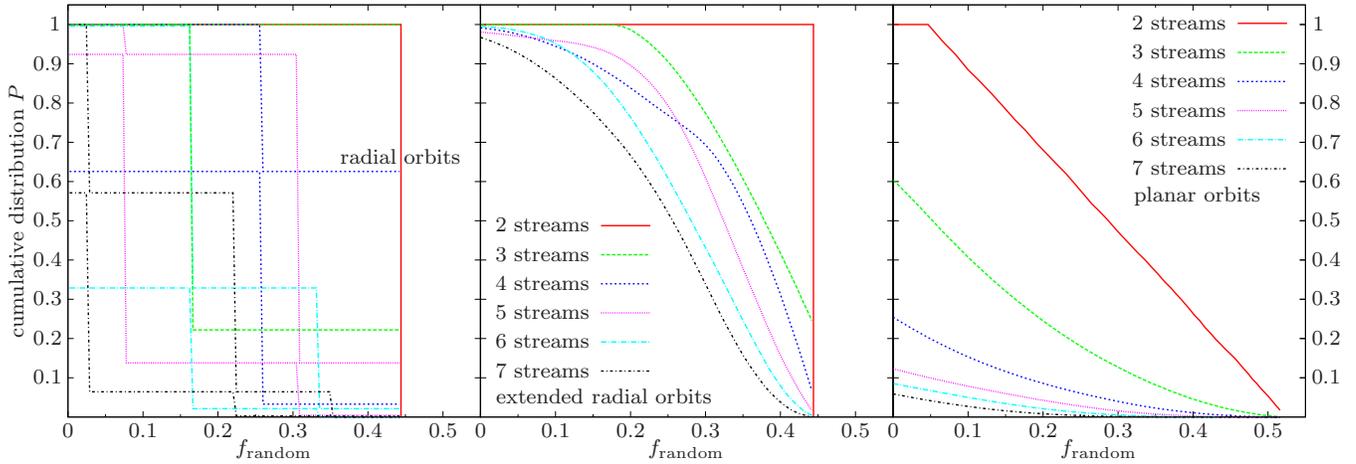}
\end{center}
\caption{Cumulative distribution $P$ of having the observed fraction of
satellites (15 out of 27 or 13 out of 27, depending on the underlying model)
within a thin plane of observed extent if one allows for a fraction
$f_{\rm random}$ of spherically, from random directions accreted satellites
calculated for various number of streams. {\it Left panel:} radial orbits
model, {\it right panel:} planar orbits model, {\it middle panel:} a new
``extended radial orbits'' model that also incorporates a varying satellite
loading factor. The model has been introduced because of the difficult
readability and limited significance of the pure radial orbits model shown in
the left panel. The panels show that a mild inflow of satellites from random
direction does not alter the conclusions drawn from either of our models.}
\label{fig:spinf}
\end{figure*}

It is a feature of our model that all satellites entering from the same stream
always stay in the same plane. It is also obvious that the stream with the
highest satellite loading factor carries always at least a fraction $1 / m$ of
the total satellites. Therefore a satellite fraction of $1 / m$ has always a
probability of $P=1.0$ regardless of the opening angle $\Psi$, as seen in
figure \ref{fig:postreams}. Generalising this one can say that the majority of
the signal is usually coming from a large value of the varying satellite
loading factor and only a minor contribution is coming from alignments of the
planes. Since in the two and in the three stream scenario, which are the most
common ones in cosmological simulations \citep{danovich} the observational data
points lie well above 50\% we can firmly state that there is nothing surprising
about the configuration seen around the Andromeda galaxy. In this model even a
four stream scenario can only just be ruled out on a one sigma level.

\citet{gerhard} present a very similar model of planar orbits, but with fixed
satellite loading fractions. They report relatively small probabilities to
reproduce the observed situation. The differences to our results arise from
their use of fixed instead of varying satellite loading fractions and from
their negligence of the two stream scenario. Both might introduce inaccuracies
since varying satellite loading fractions are clearly more physical then fixed
satellite loading fractions (compare the right panel of figure
\ref{fig:denmap}) and the two stream scenario is one of the most likely
scenarios seen in numerical simulations \citep{danovich}.

As in the radial orbits model a natural consequence of this model is that all
galaxies having at least Andromeda's mass should have such a thin plane of
satellites. Recent observations might point in this direction \citep{ibata3}.
On top of that, this model has a second natural consequence: It predicts that
high mass haloes should have several planes of satellites, namely one plane for
each stream. Indeed recent observational work \citep{tully} reports the
detection of four inclined planes in the Local Group. To make definitive
statements about multiple planes however more extended analyses are needed: One
should test the statistical significance of the planes, one should consider
velocity data or the biases introduced by the satellite survey volumes or
having associated satellites to a first plane when looking for a second one
and finally one should keep in mind that one of the planes identified is
actually offset from Andromeda's barycentre. So much more theoretical as well
as observational work is still needed in the field of multiple planes.

\subsection{Non-stream satellites}
\label{sec:addsph}

In this section we discuss the impact of satellites entering from random
directions not connected to the cold streams. Those satellites could either be
accreted at cosmic times at which at which there is no cold stream activity,
i.e. outside $z \sim 1-4$ or of sattelites not connected to any of the streams.
Satellites entering the host halo from random directions outside of any of the
streams will hardly lie in the same plane as the ones accreted as part of the
streams. Therefore extending our models to allow for satellites to enter from
random directions will significantly lower the probabilities of a certain
fraction of all satellites lying in a thin plane. Assuming that no satellite
accreted from a random direction will ever end up in the plane of the other
satellites, the fraction $f_{\rm all}$ of all the satellites however accreted
lying within a thin plane obviously decreases as $f_{\rm all} = f_{\rm stream} \
\left(1 - f_{\rm random}\right)$, where $f_{\rm stream}$ is the fraction of the
satellites accreted only through cold streams lying within a thin plane and
$f_{\rm random}$ is the fraction of spherically or randomly accreted satellites.
To calculate the probability $P_{\rm all}$ of a certain fraction of all
satellites lying in a thin plane we first fix $f_{\rm all}$ to the observed
values of 15 out of 27 or 13 out of 27, depending on the underlying model and
calculate the required fraction $f_{\rm stream}$ of the satellites accreted
through the cold streams only lying within a thin plane according to the above
formula. The probability $P_{\rm all}$ of a certain fraction of all satellites
lying in a thin plane as a function of $f_{\rm stream}$ can now be taken from our
two models. They are presented in figures \ref{fig:streams} and
\ref{fig:postreams} with $f_{\rm stream}$ indicated by different line types or by
colour, labelled as 'satellite fraction'.

In figure \ref{fig:spinf} we show how the different probabilities of having the
observed thin plane of satellites in host galaxies with a varying number of
cold streams in our two models decrease with increasing $f_{\rm random}$. In the
left panel the radial orbits model is shown. Since we have only a few discrete
data points for the satellite fraction in this model it can only show upper and
lower bounds. For configurations having two to four streams the upper bound of
the observed degree of co-planarity has a cumulative probability of more than
50\% if one allows for up to 44.4\% spherically accreted satellites. Since this
panel is difficult to read and has only limited significance we also show in
the middle panel an ``extended radial orbits model'' whose only difference to
the pure radial orbits model is that we allow for a varying satellite loading
factor. This gives narrow lines in this plot. One sees that we still get the
observed degree of co-planarity with a cumulative probability of more than 50\%
if we allow for up to 35.4\% spherically accreted satellites even in the seven
stream configuration. Scenarios with less streams can have higher fraction of
spherically accreted satellites. In the right panel we show the planar orbits
model. One sees that one still gets a thin plane of co-rotating satellites as
observed in Andromeda in 50\% of all the cases if we allow for up to 28.6\%
spherically accreted satellites in the two stream configuration or up to 4.7\%
spherically accreted satellites in the three stream configuration. So a mild
inflow of satellites from random direction does not alter the conclusions drawn
from either of our models.

\section{Conclusions}

\label{sec:conc}

Inspired by recent observations we investigated the probabilities of a certain
fraction of subhaloes lying within a thin plane around the central galaxy of
the host halo within the framework of the cold stream scenario. We performed
two sets of Monte Carlo simulations that draw randomly orientated streams
assuming that the satellites stay on circularised orbits whose planes are still
parallel to the stream. We estimated the effect of additional satellites
accreted from random directions, leading to the following results:

\begin{itemize}

\item The configuration seen around the Andromeda galaxy is a natural result of
cold stream accretion. A plane as thin as observed can be generated in both of
our models with probabilities as high as 50\% to 100\% for two or three
streams, which are the most common configurations seen in cosmological
simulations. The radial orbits model however neglects the orientations of the
rotation and only takes into account the positions of the satellites. Only the
planar orbits model takes both the positions as well as the orientations of the
rotation into account.

\item Without scattering of dwarf satellites, the Andromeda galaxy must have
been fed by more than two streams otherwise we would expect more than only 48\%
of its satellite galaxies to lie within a very thin plane with the same sens of
rotation. The most likely number of streams is three. If Andromeda’s satellite
system was produced by two streams in fact a large number of objects must have
been scattered out of the thin plane and it would be interesting to investigate
whether and how this scattering process affects the internal structure of these
satellites.

\item Since both of our models which are both based on the cold stream feeding
mechanism guarantee an initial disc that is born with a remarkable small
vertical scale-height, scattering processes as discussed by \citet{bowden} will
most likely play only a minor role.

\item A natural implication from both our models is that all galaxies having at
least the mass of Andromeda should have such a thin plane of satellites.
Interestingly recent observations point in that direction \citep{ibata3}. The
high probabilities found for the planar orbits model come about because of the
high probability to assign $\ge 48\%$ of the satellites to a single stream
which itself is a direct consequence of the fact that the most likely number of
cold streams flowing into a single host halo is relatively low, namely between
two and four \citep{danovich}. Also one should keep in mind that the radial
orbits model neglects the orientations of the rotation and only takes into
account the positions of the satellites. Only the planar orbits model takes
both the positions as well as the orientations of the rotation into account.

\item An additional implication from our second, planar orbits model is that
the satellites should naturally distribute themselves into several inclined
planes. Indeed \citet{tully} reported that there are four inclined planes
altogether in the Local Group\footnote{Concerning multiple planes one should
remember the words of caution from the end of section \ref{sec:pomresu}.}.

\item Allowing satellites to accrete from random directions lowers the
probabilities of having the required fraction of satellites in a thin disk. A
mild inflow $(\le 25\%)$ of satellites from random directions however does not
change our conclusions.

\end{itemize}

We are aware that the current analyses has its limitations. So far we do not
incorporate the evolution of the satellite's orbits, the alignment of the Milky
Way and its plane of satellites with respect to Andromeda and its respective
plane of satellites nor the possibility of having multiple planes of satellites
as observed by \citet{tully}.

We conclude that the special spatial alignment of Andromeda's satellite
galaxies can naturally be explained by cold stream accretion and simple
geometry.

\section*{Acknowledgements}
Tobias Goerdt is a Lise Meitner fellow. We thank Rodrigo Ibata, Ben Moore, Doug
Potter and Romain Teyssier for their kindness in sharing simulation and
observational data with us and Oliver Czoske for the fruitful discussions.
Tobias Goerdt would like to thank the University Observatory Munich for their
hospitality, where parts of this work were carried out. Parts of the
computational calculations were done at the Leibniz-Rechenzentrum under project
number pr86ci and at the Vienna Scientific Cluster under project number 70522.
This work was supported by FWF project number M 1590-N27 and by MINECO project
number AYA 2012-32295.

\bibliographystyle{mn2e}
\bibliography{probab18.bbl}

\label{lastpage}
\end{document}